\documentclass[10pt]{article}
\usepackage{fullpage,parskip}
\usepackage{amsmath,graphicx,hyperref}

\usepackage{amsmath}
\usepackage{xspace}
\usepackage{bbm}
\def\diag{\mathop{\rm diag}\nolimits}%

\newcommand{\clip}{\mathrm{clip}}

\newcommand{\Cc}{\mathcal{C}}
\newcommand{\Dc}{\mathcal{D}}
\newcommand{\Ec}{\mathcal{E}}

\newcommand{\Ic}{\mathcal{I}}

\newcommand{\Qc}{\mathcal{Q}}

\newcommand{\Xc}{\mathcal{X}}
\newcommand{\Yc}{\mathcal{Y}}

\newcommand{\av}{\bf a}
\newcommand{\bv}{\bf b}
\newcommand{\ev}{{\bf e}}

\newcommand{\Cv}{{\bf C}}
\newcommand{\Dv}{{\bf D}}
\newcommand{\Hv}{{\bf H}}
\newcommand{\Xv}{{\bf X}}
\newcommand{\Yv}{{\bf Y}}
\newcommand{\Zv}{{\bf Z}}

\newcommand{\xv}{{\bf x}}
\newcommand{\xvb}{\bar{{\bf x}}}
\newcommand{\yv}{{\bf y}}
\newcommand{\zv}{{\bf z}}
\newcommand{\uv}{{\bf u}}

\newcommand{\hv}{{\bf h}}
\newcommand{\sv}{{\bf s}}
\newcommand{\cv}{{\bf c}}
\newcommand{\dv}{{\bf d}}
\newcommand{\rv}{{\bf r}}

\DeclareMathOperator\E{E}
\let\P\relax
\DeclareMathOperator\P{P}

\newcommand{\Bern}{\mathrm{Bern}}

\def\textiid{i.i.d.\@\xspace}
\newcommand\iid{\ifmmode\text{ i.i.d. } \else \textiid \fi}

\newcommand{\ind}{\mathbbmss{1}}

\newcommand{\xhv}{{\bf \hat{x}}}
\newcommand{\xtv}{{\bf \tilde{x}}}

\usepackage{bbm}
\usepackage{amsthm}
\usepackage[utf8]{inputenc} 
\usepackage[T1]{fontenc}    
\usepackage{hyperref}       
\usepackage{url}            
\usepackage{booktabs}       
\usepackage{amsfonts}       
\usepackage{nicefrac}       
\usepackage{microtype}      
\usepackage{xcolor}         
\usepackage{amsmath}
\usepackage[capitalize,noabbrev]{cleveref}
\usepackage{graphicx,amssymb,comment}
\usepackage{algorithm,algorithmic}

\newtheorem{theorem}{Theorem}[section]

\newtheorem{corollary}[theorem]{Corollary}

\usepackage{verbatim}

\usepackage[utf8]{inputenc} 
\usepackage[T1]{fontenc}
\usepackage{url}
\usepackage{ifthen}
\usepackage{cite}

\title{Snapshot Compressive Imaging under Saturation: Theory, Mask Design, and Reconstruction}
\author{
    Mengyu Zhao, Shirin Jalali%
    \thanks{This article substantially extends the authors' conference
    paper presented at ICASSP 2026~\cite{zhao2025saturation}.}%
    \thanks{The authors are with the Department of Electrical and
    Computer Engineering, Rutgers University, New Brunswick, NJ, USA
    (e-mail: mengyu.zhao@rutgers.edu; shirin.jalali@rutgers.edu).}
}
\date{}

\begin{document}
%
\maketitle

\vspace{-5mm}

\begin{abstract}
Snapshot compressive imaging (SCI) acquires high-dimensional data cubes, such as videos and hyperspectral images, by optically multiplexing multiple coded frames into a single two-dimensional measurement. While this multiplexing enables high acquisition efficiency, it also increases the risk of sensor saturation: the accumulated intensity may exceed the detector dynamic range, causing clipped measurements that violate the standard linear SCI model. This paper studies SCI reconstruction under such saturated measurements from both theoretical and algorithmic perspectives. We model saturation as an element-wise clipping nonlinearity and derive a finite-sample recovery bound for compression-based SCI. The bound explicitly relates the reconstruction error to the Bernoulli mask density, the compression rate of the signal class, measurement noise, and the expected fraction of saturated measurements. The analysis reveals a principled mask-design rule: under saturation, the optimal Bernoulli mask density remains below one-half and decreases as saturation becomes stronger. Motivated by this result, we optimize mask patterns for saturated acquisition and introduce a saturation-aware plug-and-play reconstruction framework, termed \emph{Saturation-Aware PnP Net} (SAPnet), which enforces consistency with both unsaturated and clipped measurements. Experiments on standard video SCI benchmarks validate the theoretical predictions and show that SAPnet substantially improves reconstruction quality over conventional PnP-based methods, especially in strongly saturated regimes.

\end{abstract}


%
%

\section{Introduction}
\label{sec:intro}
\vspace{-1mm}

Snapshot compressive imaging (SCI) is a computational imaging paradigm that acquires high-dimensional data cubes from one or a few low-dimensional snapshots. In a typical SCI system, multiple temporal frames, spectral channels, or other data-cube slices are modulated by coded apertures and integrated by a two-dimensional sensor during a single exposure. This multiplexed acquisition strategy has enabled high-speed video imaging~\cite{llull2013coded,Hitomi11ICCV}, snapshot spectral imaging~\cite{gehm2007single,Wagadarikar08CASSI,Wagadarikar09CASSI,arce2014compressive}, and other high-throughput imaging modalities; see~\cite{yuan2021snapshot} for a comprehensive overview. By shifting part of the sensing burden from hardware acquisition to computational reconstruction, SCI reduces readout bandwidth, storage cost, and acquisition latency.

Most SCI reconstruction algorithms and theoretical analyses are based on a linear forward model. Let $\Xv_1,\ldots,\Xv_B$ denote $B$  frames and let $\Cv_1,\ldots,\Cv_B$ denote the corresponding  masks. Standard  SCI measurement systems are modeled as
\begin{equation}
    \Yv = \sum_{b=1}^{B} \Cv_b \odot \Xv_b + \Zv,\label{eq:linear-SCI}
\end{equation}
where $\Zv$ denotes measurement noise. This fundamental model has supported a broad range of model-based, plug-and-play, deep learning, and theoretical developments for SCI recovery. However, it implicitly assumes that the detector response remains linear over the entire measurement range.

In practical optical systems, this assumption can fail because of sensor saturation. Unlike conventional single-frame imaging, SCI integrates multiple coded frames or channels onto the same detector during one exposure. The accumulated intensity at a sensor pixel may therefore exceed the full-well capacity or the analog-to-digital converter range. Once this occurs, the recorded measurement is clipped at the saturation threshold and no longer follows the noisy linear model of \eqref{eq:linear-SCI}. A saturated observation is thus a nonlinear thresholded measurement that retains only partial information about the unsaturated intensity. If saturated pixels are still imposed as equality constraints, the reconstruction can be biased, with degraded contrast, missing details in bright regions, and structured artifacts.

Saturation is intrinsically coupled with optical multiplexing in SCI. A denser coded mask admits more photons and may improve the conditioning and throughput of the linear sensing operator. At the same time, it increases the integrated intensity and hence the probability of clipping. This creates a fundamental tradeoff between measurement informativeness and dynamic-range violation. The tradeoff is particularly important in video and spectral SCI, where the compression ratio, scene brightness, exposure time, mask density, and sensor dynamic range jointly determine whether the acquisition operates in the linear or saturated regime.

Although saturation and clipping have been studied in compressed sensing and computational imaging, existing results do not directly characterize SCI under saturation. Generic saturated compressed sensing theory does not capture the highly structured mask-induced SCI operator, nor the fact that mask density affects both multiplexing and clipping probability. Existing SCI theory, on the other hand, primarily assumes an unsaturated linear forward model. Consequently, there is still a lack of finite-sample recovery theory that explicitly links SCI reconstruction error to mask density and saturation severity.

This paper addresses this gap by developing a saturation-aware theory and reconstruction framework for SCI. We model sensor saturation as an element-wise clipping nonlinearity applied to the standard SCI measurement and analyze compression-based SCI recovery under this nonlinear observation model. The resulting finite-sample error bound explicitly relates the reconstruction error to the signal compression rate, measurement noise, Bernoulli mask density, and expected saturation level. The analysis yields a clear design implication: under the considered Bernoulli mask model and compression-based recovery framework, the mask density that minimizes the recovery upper bound is below one-half and decreases as saturation becomes stronger. This conclusion explains why masks designed for unsaturated SCI can become suboptimal in high-dynamic-range or high-exposure regimes.

Guided by this theoretical insight, we propose \emph{Saturation-Aware Plug-and-Play Net} (SAPnet), a reconstruction framework that distinguishes saturated and unsaturated measurements. For unsaturated pixels, SAPnet enforces standard linear measurement consistency. For saturated pixels, it imposes a clipping-aware consistency condition that respects the inequality information implied by the saturation threshold. This design preserves the flexibility of PnP SCI while incorporating the nonlinear physics of clipped measurements. Experiments on standard video SCI benchmarks validate the predicted mask-density behavior and demonstrate that SAPnet improves reconstruction accuracy and robustness over conventional PnP-based methods under different saturation levels.

\noindent \textbf{Contributions.}
The main contributions of this paper are summarized as follows:
\begin{itemize}
    \item We formulate saturated SCI as a nonlinear inverse problem by modeling sensor clipping as an element-wise saturation operator applied to the standard SCI measurement.
    \item We derive a finite-sample recovery bound for compression-based SCI under saturated measurements. The bound quantifies how reconstruction error depends on the signal compression rate, measurement noise, mask density, and saturation severity.
    \item We establish a mask-design principle for Bernoulli-coded SCI under saturation. The analysis shows that the preferred mask density is below one-half and decreases as the expected saturation level increases.
    \item We propose SAPnet, a saturation-aware PnP reconstruction framework that enforces different measurement-consistency conditions for saturated and unsaturated pixels.
    \item We validate the proposed theory and algorithm on standard video SCI benchmarks, demonstrating improved reconstruction accuracy and robustness over existing PnP-based baselines in saturated regimes.
\end{itemize}

{\bf Notation.} Throughout this paper, vectors are denoted by bold lowercase letters and matrices  by bold uppercase letters. For a matrix $\Xv\in\mathbb{R}^{n_1\times n_2}$, $\mathrm{vec}(\Xv)\in\mathbb{R}^{n}$ denotes its vectorized form, where $n=n_1n_2$. For a vector $\uv$, $u_i$ denotes its $i$th entry. Given $n\in\mathbb{N}^+$, $[n]\triangleq \{1,\ldots,n\}$. Sets are denoted by calligraphic letters, such as $\Xc$ and $\Yc$. Given set $\Xc$, $|\Xc|$ denotes its cardinality. The operator $(a)_+=\max\{a,0\}$ is applied element-wise when the argument is a vector. We use $\P(\cdot)$ and $\E[\cdot]$ to denote probability and expectation, respectively. For an index set $\Ic$, $\Pi_{\Ic}$ denotes the coordinate projection onto the entries indexed by $\Ic$.

\section{Related Work}

\subsection{SCI reconstruction}

SCI reconstruction has been extensively studied under the standard linear measurement model. Early and model-based methods exploit signal priors such as sparsity, total variation, nonlocal similarity, and low-rank structures~\cite{yuan2021snapshot}. Plug-and-play methods combine the known SCI forward model with image or video denoisers~\cite{venkatakrishnan2013pnp,chan2017pnp,reehorst2019red}, and have achieved strong performance in video and spectral SCI~\cite{ffdnet,PnP_fastdvd,Zheng20PnPCASSI}. Deep learning methods further improve reconstruction through unfolding networks~\cite{Ma19ICCV,meng2020gap,meng2023deep}, recurrent models~\cite{Cheng20ECCV_Birnat,cheng2021memory}, scalable adaptive networks~\cite{wang2021metasci}, and transformer or efficient architectures~\cite{wang2022spatial,wang2023efficientsci}. 

Most of these methods assume that the observed snapshot is an unsaturated linear projection of the latent data cube. When clipping occurs, saturated pixels are no longer exact linear measurements. Directly enforcing them as equality constraints can therefore introduce biased data consistency. SAPnet addresses this issue by incorporating the clipping model into the PnP reconstruction update.

\subsection{Optical coding and mask design}

The coded mask is a central design variable in SCI because it determines how latent frames or spectral channels are multiplexed onto the sensor. Binary random masks are widely used because of their hardware simplicity and compatibility with theoretical analysis. More structured masks, including blue-noise and grayscale coded apertures, have been explored to improve sampling behavior and reconstruction quality~\cite{correa2016spatiotemporal,diaz2019adaptive}. Recent adaptive and learned coding methods optimize the sensing process from data, including learned coded apertures and exposure patterns~\cite{vargas2021time}, deep optics for video SCI and related imaging tasks~\cite{Wang_2023_ICCV,Zhang22optics}, saliency-aware self-adaptive coding~\cite{zhao2024sasa}, and grayscale coded apertures for anti-interference spectral imaging~\cite{zhang2024anti}.

Dynamic range and sensor response have also been considered in computational imaging. Neural sensors learn programmable pixel exposures for high-dynamic-range imaging and compressive video sensing~\cite{martelneuralsensor}, deep optics methods have been developed for single-shot high-dynamic-range imaging~\cite{Metzler_2020_CVPR}, and saturation has been analyzed in Fourier multiplexed imaging~\cite{wetzstein2010sensor}. These works show that optical coding and sensor constraints are tightly coupled. However, most existing mask or optics design methods are empirical or task-driven; they do not provide a finite-sample recovery rule explaining how Bernoulli mask density should change as saturation becomes stronger.

\subsection{Theoretical analysis of SCI}

Theoretical studies of SCI aim to characterize recovery from structured snapshot measurements. Unlike classical compressed sensing with generic random projections, SCI uses sensing matrices induced by coded masks and optical integration. Snapshot compressed sensing theory therefore requires problem-specific analysis~\cite{jalali2019snapshot}. Recent work has further characterized binary mask design and the role of mask statistics under the linear SCI model~\cite{zhao2025theoretical}, and related theory has studied untrained neural networks for SCI reconstruction~\cite{zhao2024untrained}.

These results provide important foundations for SCI recovery, but they assume unsaturated linear measurements. In saturated SCI, clipped pixels provide threshold-type information rather than exact linear observations. Existing linear SCI guarantees therefore do not directly characterize reconstruction under clipping. This paper extends compression-based SCI analysis to a saturated nonlinear observation model and quantifies how the recovery error depends on mask density, compression rate, measurement noise, and saturation level.

\subsection{Saturation, clipping, and nonlinear compressed sensing}

Saturation and quantization have been studied in compressed sensing and sparse recovery. Quantized compressed sensing analyzes finite-bit measurements and the role of quantization consistency~\cite{zymnis2010compressed,jacques2011dequantizing}. One-bit compressed sensing considers the extreme nonlinear setting where only the sign of each measurement is observed~\cite{boufounos2008onebit,jacques2013robust}. More directly related to this paper, saturation-aware compressed sensing has studied clipping, saturation rejection, sparse recovery from saturated measurements, inaccurate saturated measurements, and likelihood-based recovery under Gaussian and saturation noise~\cite{laska2011,foucart2017sparse,foucart2018sparse,rencker2019sparse,banerjee2024likelihood}. These studies show that saturated measurements should be handled through nonlinear or inequality-type consistency rather than treated as ordinary noisy linear observations.

However, generic saturated compressed sensing theory does not directly apply to SCI. SCI sensing matrices are structured by coded masks rather than drawn as fully generic random projections. More importantly, mask density affects both the linear sensing operator and the clipping probability: increasing the number of open mask elements can improve photon throughput, but it also increases the accumulated intensity at the detector. Thus, saturation in SCI is simultaneously a nonlinear inverse problem and an optical design problem. Our work is tailored to this setting by providing a finite-sample recovery bound for saturated SCI, deriving a saturation-aware Bernoulli mask-density rule, and incorporating clipping-aware consistency into PnP reconstruction.
\vspace{-1mm}

\section{Problem statement}
\vspace{-1mm}

This section introduces the SCI forward model, the saturated observation model, and the compression-based recovery formulation used in our theoretical analysis. We first describe the standard linear SCI model and then extend it to account for sensor clipping.

\subsection{Linear SCI Forward Model}

Let $\Xv\in\mathbb{R}^{n_1\times n_2\times B}$ denote the target data cube, where $\Xv_b\in\mathbb{R}^{n_1\times n_2}$ is the $b$th frame or channel, for $b=1,\ldots,B$. Let $\Cv_b\in\mathbb{R}^{n_1\times n_2}$ denote the coded mask applied to $\Xv_b$. In the standard SCI model, the detector records a single two-dimensional snapshot
\begin{equation}
    \Yv
    =
    \sum_{b=1}^{B}
    \Cv_b\odot\Xv_b
    +
    \Zv,
    \label{eq:linear-SCI-matrix}
\end{equation}
where $\Zv\in\mathbb{R}^{n_1\times n_2}$ denotes additive measurement noise and $\odot$ is the Hadamard product. We can also write  \eqref{eq:linear-SCI-matrix} in a vector form. Let $\xv_b=\mathrm{vec}(\Xv_b)\in\mathbb{R}^{n}$,  $\xv=[\xv_1^\top,\ldots,\xv_B^\top]^\top\in\mathbb{R}^{nB}$, $\yv=\mathrm{vec}(\Yv)$, and 
    $\zv=\mathrm{vec}(\Zv)$. 
Also, for each mask, let $\dv_b=\mathrm{vec}(\Cv_b)\in\mathbb{R}^{n}$ and $\Dv_b=\mathrm{diag}(\dv_b)\in\mathbb{R}^{n\times n}$.
Using these definitions, the SCI sensing matrix can be written as
\begin{equation}
    \Hv=
    [\Dv_1, \Dv_2, \cdots, \Dv_B]
    \in\mathbb{R}^{n\times nB}.
    \label{eq:sci-sensing-matrix}
\end{equation}
With this notation, the linear SCI model becomes
\begin{equation}
    \yv
    =
    \Hv\xv+\zv.
    \label{eq:linear-SCI-vector}
\end{equation}
The goal of SCI reconstruction is to estimate the high-dimensional signal $\xv\in\mathbb{R}^{nB}$ from the low-dimensional measurement $\yv\in\mathbb{R}^{n}$, given the sensing matrix $\Hv$. Since $B>1$ in snapshot compression, this inverse problem is underdetermined and must be regularized by structural assumptions on the signal class.

In the theoretical analysis, we focus on binary Bernoulli masks. Specifically, we assume for $b=1,\ldots,B$, and $i=1,\ldots,n$, $\{d_{b,i}\}$ are independently and identically distributed (i.i.d.)  as $\Bern(p)$. Here,
 $p\in(0,1)$ denotes the mask density controlling the number of active coded components contributing to each sensor pixel. Under saturation, it also controls the probability that the accumulated intensity exceeds the detector dynamic range.

\subsection{Saturated SCI Observation Model}

The linear model \eqref{eq:linear-SCI-vector} assumes that the detector response remains within its dynamic range. In practice, this assumption may fail because SCI integrates multiple coded frames or channels during a single exposure. When the accumulated intensity exceeds the full-well capacity or the analog-to-digital converter range, the recorded value is clipped at a saturation threshold.

Let $\yv= \Hv\xv+\zv$ denote the  unsaturated noisy measurement vector. For a threshold $T>0$, the observed saturated measurement is modeled as
\begin{equation}
    \yv_T
    =
    \clip(\yv;T),
    \label{eq:saturated-measurement}
\end{equation}
where the clipping operator is applied element-wise as follows. The index set of saturated  measurements is defined as 
\begin{equation}
    \Ic_s
    =
    \{j\in[n]:(\Hv\xv)_i=\sum_{i=1}^BD_{ij}x_{ij} \geq T\}.
    \label{eq:saturated-index-set}
\end{equation}
Then, the saturated measurement $y_{T,i}$ is defined as
\begin{align}
    y_{T,i}=\left\{\begin{array}{ll}
        T, &\quad \text{if}\;i\in\Ic_s  \\
        \textcolor{white}{0}\\
         y_i,&\quad \text{if}\; i\in\Ic_s^c. 
    \end{array}\right.
\end{align}

Under this model,  a saturated measurement does not specify the exact value of ${y}_i$; it only provides the inequality information ${y}_i\geq T$. This distinction is central to the recovery formulation below.

It is useful to identify the unsaturated limiting case. Suppose that the signal is nonnegative and bounded as
\begin{equation}
    0\leq x_{b,i}\leq \rho/2,
    \qquad
    b=1,\ldots,B,\quad i=1,\ldots,n.
    \label{eq:signal-bound}
\end{equation}
For binary masks and zero noise, we have
\begin{equation}
    (\Hv\xv)_i
    =
    \sum_{b=1}^{B}d_{b,i}x_{b,i}
    \leq
    \frac{B\rho}{2}.
\end{equation}
Therefore, if $T\geq B\rho/2$, then $\yv_T=\Hv\xv+\zv$ for all admissible signals, and no signal-induced saturation occurs. Therefore,  the main regime of interest is $T<B\rho/2$, where saturation can occur depending on the scene intensity, mask density, compression ratio and exposure time.

We also define the empirical saturation fraction as
\begin{equation}
    \alpha_T
    =
    \frac{|\Ic_s|}{n}.
    \label{eq:saturation-fraction}
\end{equation}
In other words, $\alpha_T$ denotes the fraction of measurements that are saturated. This is random number that depends on masks, $\xv$ and $T$. 
Given $\xv\in\mathbb{R}^{nB}$, let $ p_s(\xv;T)$ denote the expected value of $  \alpha_T$, i.e., the expected fraction of saturated measurements for a given input $\xv$ and threshold $T$. More precisely, 
\begin{equation}
    p_s(\xv;T)
    \triangleq
    \E[\alpha_T]
    =
    \frac{1}{n}
    \sum_{i=1}^{n}
    \P((\Hv\xv)_i\geq T),
    \label{eq:saturation-probability}
\end{equation}
where the probability is taken over the random masks.  The dependence on the Bernoulli mask density $p$ is implicit in \eqref{eq:saturation-probability}. This quantity characterizes the effect of saturation on SCI recovery. As we show in Section \ref{sec:compression-based-theory}, $p_s(\xv;T)$ affects the performance of compression-based recovery described in \eqref{eq:CSP-saturated-projection}.

\subsection{Compression-Based Recovery under Saturation}\label{sec:compression-based-theory}

We adopt the compression-based recovery framework commonly used in SCI theory~\cite{jalali2019snapshot,zhao2025theoretical}. Let $\Qc\subset\mathbb{R}^{nB}$ denote a compact signal class. A lossy compression code consists of an encoder $f$ and a decoder $g$, and induces a finite codebook
\begin{equation}
    \Cc
    =
    \{g(f(\xv)):\xv\in\Qc\}
    \subset\mathbb{R}^{nB}.
    \label{eq:codebook}
\end{equation}
For a rate-$r$ code measured in bits per signal entry, the codebook satisfies
\begin{equation}
    |\Cc|
    \leq
    2^{nBr}.
    \label{eq:codebook-size}
\end{equation}
The corresponding worst-case normalized distortion is
\begin{equation}
    \delta
    =
    \sup_{\xv\in\Qc}
    \frac{1}{nB}
    \|\xv-g(f(\xv))\|_2^2.
    \label{eq:compression-distortion}
\end{equation}
The codebook $\Cc$ provides a finite approximation of the signal class $\Qc$, while $\delta$ quantifies the approximation error.

In the unsaturated linear setting, compression-based signal pursuit (CSP) selects the codeword whose SCI measurement best matches the observation:
\begin{equation}
    \hat{\xv}_{\mathrm{lin}}
    \in
    \arg\min_{\cv\in\Cc}
    \|\yv-\Hv\cv\|_2^2.
    \label{eq:CSP-linear}
\end{equation}
This estimator is primarily used as an analytical tool for deriving finite-sample recovery guarantees under structured signal models.

In the saturated setting, the observation is $\yv_T$ rather than $\yv$. Directly substituting $\yv_T$ into \eqref{eq:CSP-linear} would incorrectly enforce $(\Hv\cv)_i=T$ at saturated pixels. This is physically inaccurate because a saturated pixel only implies that the corresponding latent unsaturated measurement is no smaller than the threshold. To encode this inequality information, we consider the following saturation-aware CSP estimator:
\begin{equation}
\begin{aligned}
    \hat{\xv}
    \in
    \arg\min_{\cv\in\Cc}
    \Bigg\{
    &
    \sum_{i\in\Ic_s^c}
    \big(
    y_{T,i}-(\Hv\cv)_i
    \big)^2
    \\
    &+
    \sum_{i\in\Ic_s}
    \big(
    T-(\Hv\cv)_i
    \big)_+^2
    \Bigg\}.
\end{aligned}
\label{eq:CSP-T}
\end{equation}
The first term imposes standard equality consistency on unsaturated measurements. The second term is a squared hinge penalty for saturated measurements: it penalizes a candidate codeword only when its predicted measurement falls below the saturation threshold. If $(\Hv\cv)_i\geq T$ for a saturated pixel, the candidate is consistent with the clipping event and incurs no penalty at that location.

Equivalently, \eqref{eq:CSP-T} can be written in projection form as
\begin{equation}
\begin{aligned}
    \hat{\xv}
    \in
    \arg\min_{\cv\in\Cc}
    &
    \left\|
    \Pi_{\Ic_s^c}
    (\Hv\cv-\yv_T)
    \right\|_2^2
    \\
    &+
    \left\|
    \left(
    T\mathbf{1}_{|\Ic_s|}
    -
    \Pi_{\Ic_s}\Hv\cv
    \right)_+
    \right\|_2^2,
\end{aligned}
\label{eq:CSP-saturated-projection}
\end{equation}
where $\mathbf{1}_{|\Ic_s|}$ is the all-one vector in $\mathbb{R}^{|\Ic_s|}$. This formulation separates the two types of information available under clipping: equality information from unsaturated pixels and inequality information from saturated pixels.

The central question is how the nonlinear observation model \eqref{eq:saturated-measurement} affects SCI recovery. In the next section, we analyze \eqref{eq:CSP-T} and derive finite-sample recovery guarantees. The resulting bound quantifies the impact of clipping through the saturation level $p_s(\xv,\zv;T)$ and reveals how the Bernoulli mask density $p$ should be chosen to mitigate saturation-induced reconstruction error.

\section{Theoretical Analysis}
\label{sec:theory}
We now analyze the saturation-aware CSP estimator in
\eqref{eq:CSP-T} under the clipped SCI observation model in
\eqref{eq:saturated-measurement}. The signal class $\Qc$, codebook $\Cc$,
compression distortion $\delta$, and all measurement-model quantities are
those introduced in Section~III. Following
\eqref{eq:saturation-probability}, all probability statements are with
respect to the random Bernoulli masks, with the noise realization held
fixed. 

The following theorem quantifies how this expected saturation fraction
enters the recovery bound for compression-based SCI.

\begin{theorem} 
\label{thm:1-sat} Consider $\Qc\subset\mathbb{R}^{nB}$, and assume that $\|\xv\|_{\infty}\leq\rho/2$ for every $\xv\in\Qc$. For $\xv\in\Qc$, consider the saturated observation model in \eqref{eq:saturated-measurement}. Assume that $\|\zv\|_2\leq\epsilon_z$ for some $\epsilon_z\geq0$. Further, assume that the diagonal entries of $\Dv_1,\ldots,\Dv_B$ are i.i.d.\ ${\rm Bern}(p)$, where $p\in(0,1)$. Let $\xhv$ be the solution of \eqref{eq:CSP-T}, and define \begin{equation} \beta_T = \left(\frac{B\rho}{2}-T\right)^+. 
\label{eq:beta-T} 
\end{equation} 

Then, for any $\epsilon_1,\epsilon_2>0$, 
\begin{align} 
&\frac{1}{\sqrt{nB}}\|\xv-\xhv\|_2 \leq{}
 \sqrt{\left(1+\frac{Bp}{1-p}\right)\delta} + 2\rho\sqrt{\frac{\epsilon_1}{2p(1-p)}} \nonumber\\ 
&+ 2\sqrt{ \frac{ \big(p_s(\xv;T)+\epsilon_2\big) \beta_T(\beta_T+2B\rho) }{ p(1-p)B } } + \frac{2\epsilon_z}{\sqrt{p(1-p)nB}}. 
\label{eq:thm1-sat} 
\end{align} 
with probability at least \begin{equation} 
1 - 2^{Br+1} \exp\left( -\frac{n\epsilon_1^2}{2B^2} \right) - \exp(-2n\epsilon_2^2). 
\label{eq:thm1-sat-probability} 
\end{equation} 
\end{theorem}

The four terms in \eqref{eq:thm1-sat} correspond to the compression distortion, mask-concentration error, saturation-induced error, and additive noise, respectively. In particular, the saturation term depends jointly on the expected saturation fraction $p_s(\xv;T)$ and the clipping gap $\beta_T$. 

In the no-clipping and noiseless regime, i.e., $p_s(\xv;T)=0$, $\beta_T=0$, and $\epsilon_z=0$, the saturation and noise terms vanish, leaving the same compression-distortion and mask-concentration terms as in the saturation-free CSP analysis of \cite{zhao2025theoretical}. Moreover, both $p_s(\xv;T)$ and $\beta_T$ are nonincreasing functions of $T$, so the saturation contribution decreases as the clipping threshold increases and vanishes whenever $\beta_T=0$. 

The recovery bound also provides a mask-design rule. For $T>0$, let $p_T^*$ denote a value of $p$ that minimizes the upper bound in Theorem~\ref{thm:1-sat}, where the dependence of $p_s(\xv;T)$ on $p$ is implicit.

\begin{corollary}
\label{cor:1}
Consider the same setting as in Theorem~\ref{thm:1-sat}, and assume that
$\epsilon_z=0$. For every $T>0$, $p_T^*<1/2$. Furthermore, $p_T^*$ is
an nondecreasing function of $T$.
\end{corollary}

Corollary~\ref{cor:1} states that the bound-optimal Bernoulli mask
density remains below one half, as in the saturation-free setting
\cite{zhao2025theoretical}. It also states that stronger saturation calls
for a smaller fraction of nonzero mask entries. Thus, as the clipping
threshold decreases, the mask density that minimizes the recovery bound
also decreases.

The key term in Theorem \ref{thm:1-sat} that is specific to the case of saturated measurements is $p_s(\xv;T)$, defined in \eqref{eq:saturation-probability}. $p_s(\xv;T)$ characterizes the expected  percentage of saturated measurements and is directly connected to the properties of $\xv$.


To shed some light on the connection of $p_s(\xv;T)$ and properties of $\xv$, 
define the saturation indicator at location $j\in\{1,\ldots,n\}$, as 
\[
    S_j
    \triangleq
    \ind_{\left\{
        \sum_{i=1}^{B}D_{ij}x_{ij}
        \geq T
    \right\}}.
\]
Using this definition the number of saturated measurements can be written as $|\Ic_s|=\sum_{j=1}^nS_j$. Since the masks' elements  $\{D_{ij}\}_{(i,j)\in [B]\times [n]} $are i.i.d.~as $\Bern(p)$,
$S_1,\ldots,S_n$ are independent Bernoulli random variables with generally different means. Therefore, as shown in the proof of Theorem \ref{thm:1-sat}, ${1\over n}|\Ic_s|$ concentrates around its mean, which is $p_s(\xv;T)$. 

We next introduce a mean-frame approximation to establish  a more explicit
connection  between $p_s(\xv;T)$ and the properties of $\xv$. Define the temporal mean frame of $\xv$ as
\[
    \xvb
    \triangleq
    \frac{1}{B}
    \sum_{i=1}^{B}\xv_i.
\]
Note that 
\begin{align}
    \Hv\xv
    &=
    \sum_{i=1}^{B}\Dv_i\xv_i
    \nonumber\\
    &=
    \Big(
        \sum_{i=1}^{B}\Dv_i
    \Big)\xvb
    +
    \sum_{i=1}^{B}
    \Dv_i(\xv_i-\xvb).
    \label{eq:mean-frame-decomposition}
\end{align}
The first term is determined by the temporal mean, whereas the second
term captures frame-to-frame deviations. For a fixed input $\xv$, note that  $\E[\sum_{i=1}^{B}\Dv_i]=BpI_n$ and 
\[
\E[\sum_{i=1}^{B}
    \Dv_i(\xv_i-\xvb)]=p\sum_{i=1}^{B}
    (\xv_i-\xvb)={\bf 0}_{n}.
\]
Thus, the second term in \eqref{eq:mean-frame-decomposition} is a zero-mean fluctuation around the contribution determined by the temporal mean frame. Its effect on saturation depends on the magnitude of the frame-to-frame deviations. In particular, when these deviations are small, the first term, $(
        \sum_{i=1}^{B}\Dv_i
    )\xvb$
provides a useful approximation to the measurement for interpreting the saturation behavior. Consequently, for nonnegative signals, larger mean-frame intensities are generally associated with a larger expected fraction of saturated measurements. This relationship is further examined empirically in Section~ \ref{subsec:sat_statistics}.

\vspace{0.2cm}


\section{Saturation-Aware PnP Reconstruction}
\label{sec:sapnet}
The saturation-aware CSP estimator in \eqref{eq:CSP-T} enforces different consistency conditions for unsaturated and saturated measurements. However, directly solving \eqref{eq:CSP-T}  is infeasible, primarily  due to the exponential size of the codebook. Similar computational complexity issues arise when solving the CSP optimization proposed for SCI without saturation \cite{jalali2019snapshot}. One approach to overcome this problem and develop an efficient solution for approximating the solution of such optimizations is to employ projected gradient descent (PGD). In \cite{jalali2019snapshot}, both PGD and its variant generalized alternating projection (GAP),  were studied and shown, both theoretically and algorithmically,   to  recover the solution. Inspired by the success to such methods for unsaturated  measurements, we propose SAPnet, which combines the generalized alternating projection (GAP) update corresponding to the objective function in \eqref{eq:CSP-T} with  pretrained FastDVDnet denoiser~\cite{PnP_fastdvd} for projection.


At iteration $t$, SAPnet first computes the predicted measurement $\rv^t=\Hv\xv^t$. On the unsaturated set, it uses the standard linear residual $\yv_T-\rv^t$. On the saturated set, the clipping event only requires the predicted pre-clipping measurement to be at least $T$. SAPnet therefore applies the one-sided residual $(T\mathbf{1}-\rv^t)_+$: a prediction below $T$ is corrected upward, whereas a prediction at or above $T$ receives no equality correction. The two residuals are combined and used in a GAP data-consistency step, followed by video denoising. Algorithm~\ref{algo:sapnet} summarizes the complete procedure.

\begin{algorithm}[t]
\caption{SAPnet for Saturated SCI Reconstruction}
\label{algo:sapnet}
\begin{algorithmic}[1]
\REQUIRE Sensing matrix $\Hv$; clipped measurement $\yv_T$;
threshold $T$; stepsize $\mu$; denoiser $\Dc$; number of
iterations $K$.
\STATE Initialize $\xv^{0}=\mathbf{0}$.
\STATE Set $\bv_u=\mathbf{1}_{\{\yv_T<T\}}$ and
$\bv_s=\mathbf{1}_{\{\yv_T=T\}}$.
\STATE Set $\av=\diag(\Hv\Hv^\top)$.
\FOR{$t=0$ \textbf{to} $K-1$}
    \STATE Forward projection:
    $\rv^{t}=\Hv\xv^{t}$.
    \STATE Unsaturated residual:
    $\ev_u^{t}=\bv_u\odot(\yv_T-\rv^{t})$.
    \STATE Saturated residual:
    $\ev_s^{t}=\bv_s\odot(T\mathbf{1}-\rv^{t})_+$.
    \STATE Combined residual:
    $\widetilde{\ev}^{t}=\ev_u^{t}+\ev_s^{t}$.
    \STATE GAP normalization:
    $\bar{e}_j^{t}=\widetilde{e}_j^{t}/a_j$ for all $j$.
    \STATE GAP update:
    $\sv^{t+1}
    =
    \xv^{t}
    +
    \mu\Hv^\top\bar{\ev}^{t}$.
    \STATE Denoising update:
    $\xv^{t+1}=\Dc(\sv^{t+1})$.
\ENDFOR
\STATE \textbf{Output:} $\hat{\xv}=\xv^{K}$.
\end{algorithmic}
\end{algorithm}

For SCI,
\[
    \Hv\Hv^\top
    =
    \sum_{b=1}^{B}\Dv_b^2
\]
is diagonal, so the GAP normalization requires only element-wise division rather than a full matrix inversion. If $a_j=0$, the quotient in Algorithm~\ref{algo:sapnet} is interpreted in the Moore--Penrose sense, namely $\bar e_j^t=0$. This convention covers locations for which all $B$ mask entries are zero.

Each SAPnet iteration requires one application of $\Hv$ and one application of $\Hv^\top$, both of which have complexity $\mathcal{O}(nB)$ for the SCI sensing operator. The residual construction, masking, and GAP normalization require only $\mathcal{O}(n)$ additional operations and $\mathcal{O}(n)$ auxiliary storage. Therefore, excluding the denoiser, the total complexity over $K$ iterations is $\mathcal{O}(KnB)$, matching the order of standard PnP-GAP. In practice, the FastDVDnet denoising step remains the main computational component, while the saturation-aware modification adds only element-wise operations and no trainable parameters. Consequently, it requires no saturation-specific retraining. When no measurement is saturated, $\bv_s=\mathbf{0}$ and $\bv_u=\mathbf{1}$, so SAPnet reduces to the standard PnP-GAP iteration. The recovery guarantee in Theorem~\ref{thm:1-sat} applies to the finite-codebook estimator in \eqref{eq:CSP-T}; SAPnet is its practical PnP counterpart and is not directly covered by that guarantee.

\section{Experiments}

This section evaluates both the theoretical findings for the compression-based estimator and the practical SAPnet method introduced in Section~\ref{sec:sapnet}. We examine four questions: (i) how the empirical saturation fraction varies with the clipping threshold and scene content; (ii) whether clipping-aware data consistency improves reconstruction over standard PnP-FastDVDnet; (iii) whether the empirically preferred Bernoulli mask density qualitatively follows the bound-based rule in Corollary~\ref{cor:1}; and (iv) whether the improvement persists in the presence of pre-clipping noise.

\subsection{Experimental Protocol}
\label{subsec:exp_protocol}

\textbf{Datasets.}
We evaluate on six standard grayscale video SCI benchmark sequences: \texttt{Kobe}, \texttt{Runner}, \texttt{Drop}, \texttt{Traffic}, \texttt{Aerial}, and \texttt{Crash}~\cite{PnP_fastdvd}. Each video cube has spatial resolution $256\times256$ and compression ratio $B=8$. The video frames are normalized before measurement simulation.

\textbf{Measurement generation.}
For each ground-truth video cube, we sample binary SCI masks from $\Bern(p)$ and generate measurements according to \eqref{eq:saturated-measurement}. Unless otherwise specified, we use the noiseless setting $\zv=\mathbf{0}$, which satisfies the noise condition in Theorem~\ref{thm:1-sat}. Since the normalized benchmark videos do not provide physical radiance, exposure, or full-well-capacity information, we report the clipping threshold through the normalized ratio $T/B$. Within this synthetic setting, a smaller $T/B$ represents stronger clipping relative to the measurement scale and produces a larger fraction of saturated measurements. Physically, this change can model an increase in exposure relative to the sensor saturation level or a decrease in sensor dynamic range.

\textbf{Baselines and metrics.}
We compare SAPnet with PnP-FastDVDnet~\cite{PnP_fastdvd}, a strong PnP baseline for video SCI reconstruction. Both methods use the same measurements and the same masks for each experimental setting. Unless otherwise stated, the default mask density is $p=0.5$. Reconstruction quality is measured by peak signal-to-noise ratio (PSNR). For the mask-density study, we sweep
\[
    p\in\{0.1,0.2,\ldots,0.9\}.
\]
All reported comparisons are performed under matched sensing conditions so that the observed differences are attributable to the reconstruction update or the mask density.

\subsection{Saturation Statistics}
\label{subsec:sat_statistics}

We first quantify how frequently the SCI measurements are clipped. For a fixed video cube, noise vector, and mask realization, the empirical saturation fraction is
\begin{equation}
    \alpha_T
    =
    \frac{|\Ic_s|}{n},
    \qquad
    \Ic_s=
    \{j:\bar y_j\geq T\}.
    \label{eq:emp_sat_fraction}
\end{equation}
Averaging $\alpha_T$ over independent mask realizations provides a Monte Carlo estimate of the theoretical saturation fraction
$p_s(\xv;T)$ defined in \eqref{eq:saturation-probability}.

\begin{figure}[t]
    \centering
    \includegraphics[width=\linewidth]{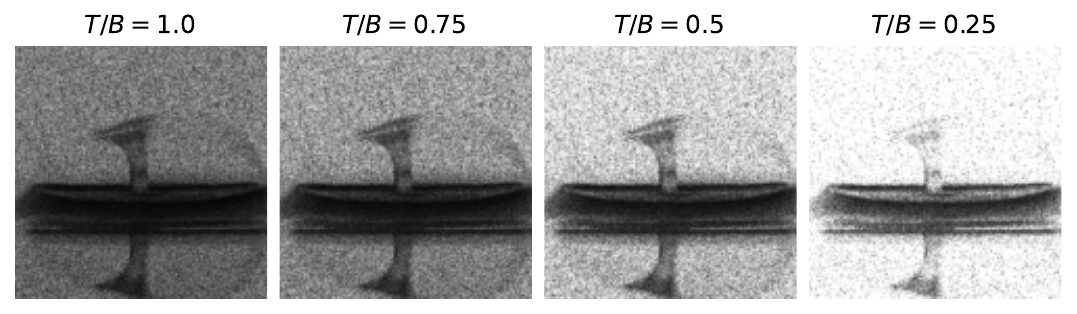}
    \includegraphics[width=\linewidth]{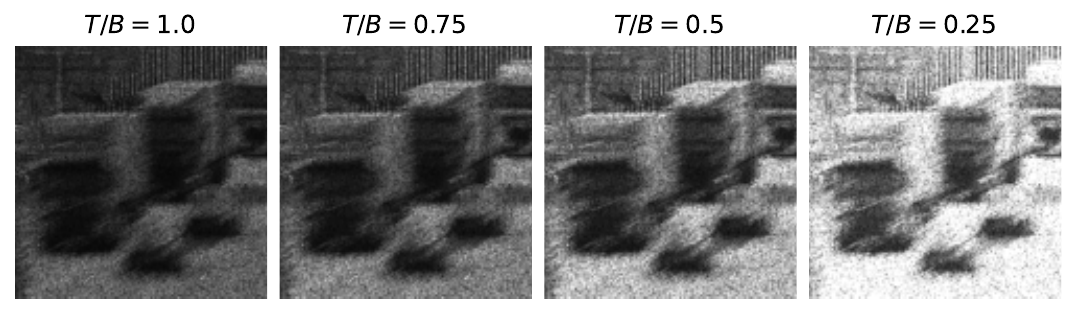}
    \caption{Clipped SCI measurements of \texttt{Drop} and \texttt{Traffic} under different saturation thresholds. Smaller $T$ produces stronger clipping and removes more high-intensity measurement information.}
    \label{fig:sat_meas}
\end{figure}

\begin{figure}[t]
    \centering
    \includegraphics[width=\linewidth]{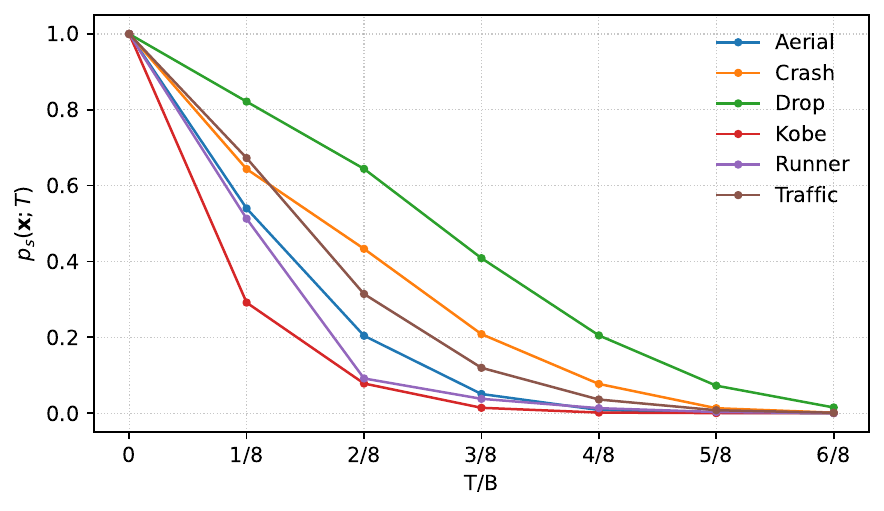}
    \caption{Empirical saturation fraction $\alpha_T$ for different video inputs as a function of the normalized threshold $T/B$. Masks are sampled from $\Bern(0.5)$.}
    \label{fig:sat_prob}
\end{figure}

Fig.~\ref{fig:sat_meas} visualizes the spatial effect of saturation for \texttt{Drop} and \texttt{Traffic}. As $T$ decreases, progressively more measurements reach the clipping threshold, enlarging the saturated regions and removing the relative variations among high-intensity measurements.

Fig.~\ref{fig:sat_prob} shows estimated $\alpha_T$ corresponding to different videos. It can be observed that, as argued in Section \ref{sec:theory}, data frames with higher  temporal means have larger percentage of saturated measurements. In other words,  at a location with a larger temporal mean intensity, fewer active Bernoulli mask elements are needed for the accumulated measurement to exceed $T$. Therefore, videos containing a larger fraction of persistently bright regions tend to exhibit larger saturation fractions, even when the mask density and normalized threshold are identical. The higher curves of \texttt{Drop} and \texttt{Crash}, compared with \texttt{Kobe}, are consistent with this interpretation. The temporal deviations from the mean frame and the fixed noise realization further affect the exact saturation probability and account for behavior not captured by the mean-frame approximation.

The recovery bound in Theorem~\ref{thm:1-sat} depends explicitly on the saturation probability $p_s(\xv;T)$. The empirical results above show that this quantity varies with both the saturation threshold and the underlying video content, suggesting that the impact of saturation on reconstruction may likewise vary across different settings. This motivates the reconstruction experiments in the following subsections, where we examine how different saturation levels affect recovery performance.

\subsection{Reconstruction under Saturation}
\label{subsec:reco_saturation}

Table~\ref{tab:sap_gain} reports the PSNR improvement of SAPnet over PnP-FastDVDnet under three saturation levels using the default mask density $p=0.5$. The gain is largest when saturation is severe. At $T/B=0.25$, SAPnet improves the average PSNR by $11.94$ dB over the six benchmark sequences. At $T/B=0.5$, the average improvement is $5.58$ dB. When the threshold increases to $T/B=0.75$, saturation becomes weak for most scenes, and the average improvement decreases to $1.11$ dB. This trend agrees with the theory: As $T$ increases, fewer measurements are assigned to the saturated mask $\bv_s$. SAPnet therefore applies its one-sided correction less frequently and approaches the standard PnP-GAP update, explaining the smaller empirical gain. Separately, the decrease of $p_s(\xv;T)$ and $\beta_T$ is consistent with the smaller saturation-dependent term in Theorem~\ref{thm:1-sat}.

\begin{table}[t]
\centering
\caption{PSNR improvement of SAPnet over PnP-FastDVDnet under different saturation thresholds using masks sampled from $\Bern(0.5)$.}
\label{tab:sap_gain}
\begin{tabular}{@{}lccc@{}}
\toprule
\textbf{Dataset}
& \textbf{$T/B=0.25$}
& \textbf{$T/B=0.5$}
& \textbf{$T/B=0.75$} \\
\midrule
\texttt{Runner}  & +13.148 & +2.497  & +0.019 \\
\texttt{Drop}    & +30.017 & +24.677 & +6.594 \\
\texttt{Crash}   & +11.189 & +1.940  & -0.028 \\
\texttt{Traffic} & +6.352  & +3.259  & +0.072 \\
\texttt{Kobe}    & +6.604  & +0.430  & -0.009 \\
\texttt{Aerial}  & +4.325  & +0.690  & +0.017 \\
\midrule
\textbf{Average} & +11.939 & +5.582 & +1.111 \\
\bottomrule
\end{tabular}
\end{table}

The largest gains occur on \texttt{Drop}, which also exhibits one of the largest saturation fractions in Fig.~\ref{fig:sat_prob}. In contrast, when saturation is weak, SAPnet and PnP-FastDVDnet become nearly identical because most measurements are governed by the same linear consistency condition. The small negative gains observed in a few weak-saturation cases are negligible and indicate that the saturation-aware update does not substantially perturb the reconstruction when clipping is almost absent.

\subsection{Effect of Bernoulli Mask Density}
\label{subsec:mask_density}

\begin{figure}[t]
    \centering
    \includegraphics[width=\linewidth]{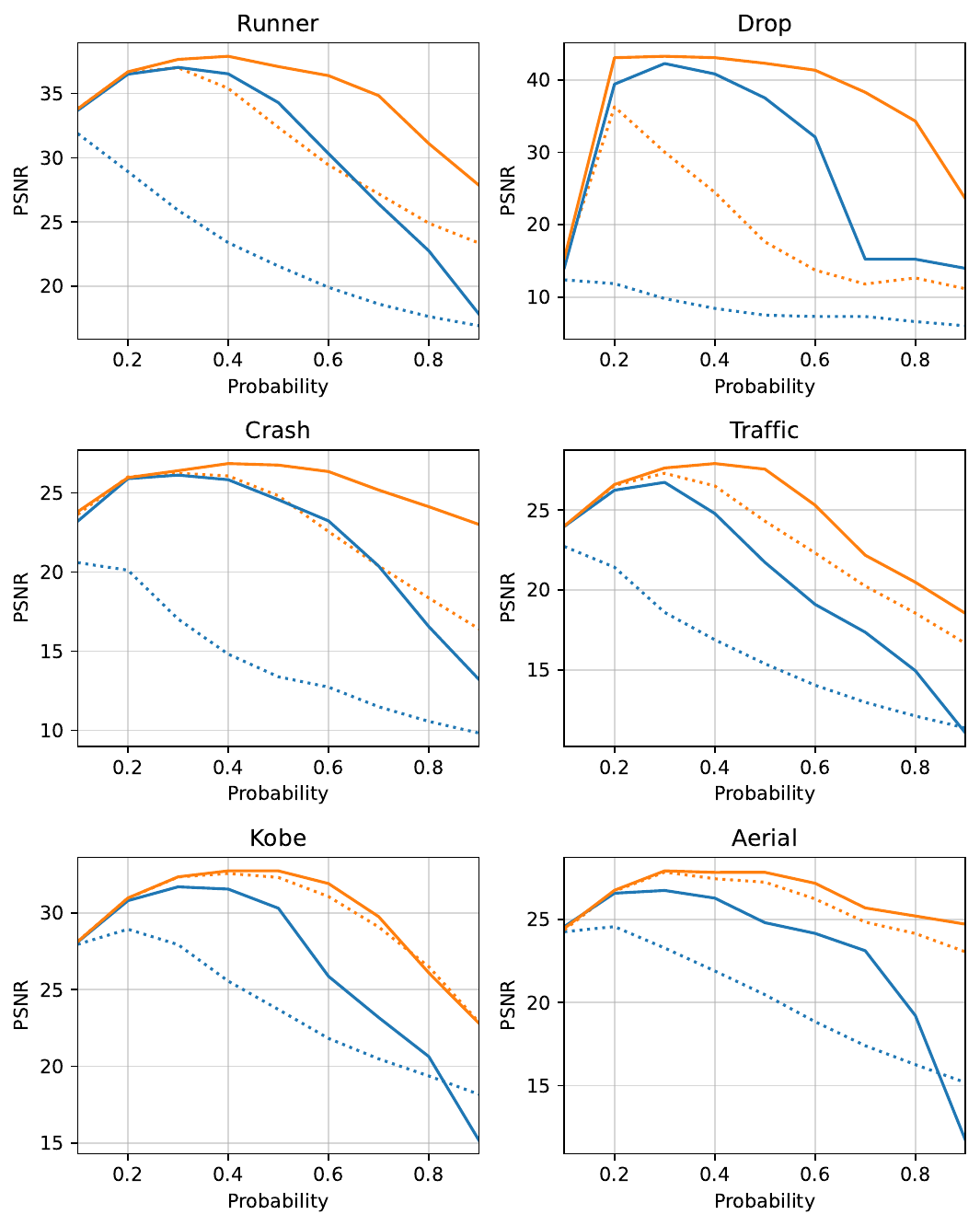}
    \caption{Reconstruction PSNR under different Bernoulli mask densities $p$. Solid curves correspond to SAPnet, and dotted curves correspond to PnP-FastDVDnet. The two threshold settings are $T/B=0.25$ and $T/B=0.5$.}
    \label{fig:sat_reco}
\end{figure}

We next evaluate the mask-density prediction in Corollary~\ref{cor:1}. For each benchmark sequence and saturation threshold, we sample binary masks from $\Bern(p)$ with
\[
    p\in\{0.1,0.2,\ldots,0.9\}.
\]
Fig.~\ref{fig:sat_reco} compares SAPnet and PnP-FastDVDnet across these mask densities.

The empirical maximizers occur below $p=0.5$ in saturated regimes, consistent with Corollary~\ref{cor:1}. Moreover, the preferred density shifts to smaller values as saturation becomes stronger. This behavior reflects the tradeoff identified by the theory. Increasing $p$ improves photon throughput and can improve the conditioning of the unsaturated linear SCI measurement, but it also increases the accumulated intensity at each sensor pixel and therefore raises the clipping probability. Under severe saturation, sparser masks reduce clipping and yield better reconstruction.

The comparison between SAPnet and PnP-FastDVDnet also shows that algorithmic saturation awareness and mask selection are complementary. SAPnet improves reconstruction for a fixed mask by enforcing clipping-consistent updates, while reducing the mask density mitigates saturation at the acquisition stage.

\subsection{Robustness to Pre-Clipping Noise}
\label{subsec:noisy_experiment}

Finally, we evaluate reconstruction when additive noise is present before clipping. Measurements are generated as
\begin{equation}
    \yv_T=\clip(\Hv\xv+\zv;T),
\end{equation}
where $\zv$ is additive Gaussian noise. In the reported experiment, the noise standard deviation is $\sigma=10$ on the 8-bit intensity scale, corresponding approximately to $0.04$ after normalization.

\begin{figure}[t]
    \centering
    \includegraphics[width=\linewidth]{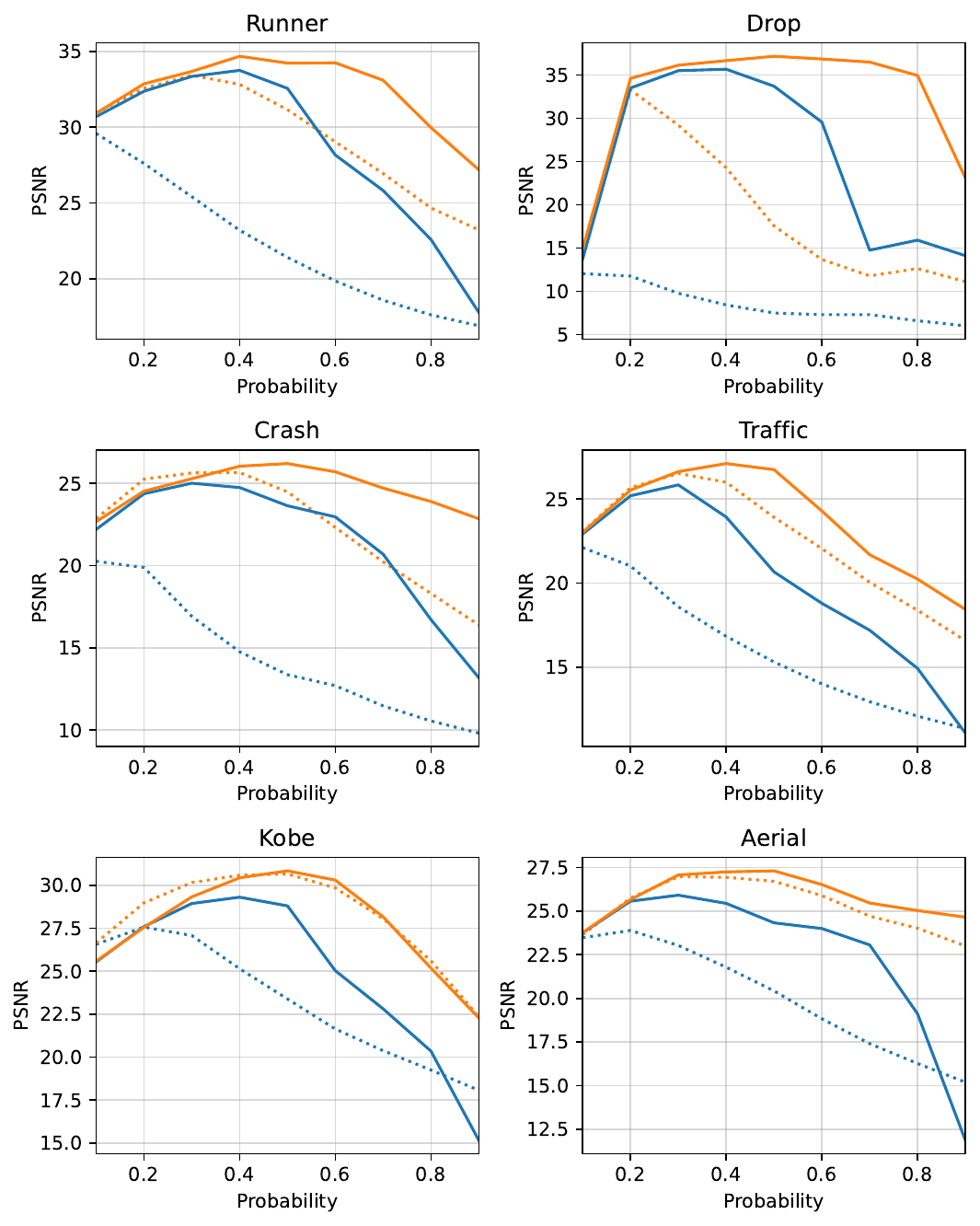}
    \caption{Noisy-measurement reconstruction PSNR under different Bernoulli mask densities $p$. Solid curves correspond to SAPnet, and dotted curves correspond to PnP-FastDVDnet. The two threshold settings are $T/B=0.25$ and $T/B=0.5$.}
    \label{fig:sat_reco_noise}
\end{figure}

Fig.~\ref{fig:sat_reco_noise} shows that SAPnet remains robust when noise is added before clipping and continues to improve over the baseline under saturation. Compared with the noiseless case, the empirically preferred mask density can shift slightly because noise and saturation impose competing requirements. Denser masks may improve the effective measurement signal-to-noise ratio when clipping is moderate, whereas sparser masks remain preferable when clipping dominates. This qualitative behavior is consistent with Theorem~\ref{thm:1-sat}, where the bound contains both a noise-dependent term and a saturation-dependent term.

\subsection{Discussion}
\label{subsec:exp_discussion}

The experimental results support the main theoretical predictions. First, the severity of saturated SCI reconstruction is governed by the fraction of clipped measurements, which varies significantly across scenes and thresholds. Second, treating saturated measurements as inequality-type constraints improves reconstruction compared with imposing them as exact linear measurements. Third, the best Bernoulli mask density in saturated regimes is below $0.5$ and decreases as clipping becomes stronger. These findings indicate that saturated SCI should be addressed jointly at the algorithmic and optical-coding levels: the reconstruction method should respect the clipping model, and the mask density should be selected according to the expected saturation level.




\section{Proofs of Main Theory}\label{proof-saturated-csp}
In this section we present the proofs of Theorem~\ref{thm:1-sat} and
Corollary~\ref{cor:1}.

\subsection{Proof of Theorem~\ref{thm:1-sat}}

Let $\xtv=g(f(\xv))\in\Cc$ be the codeword associated with $\xv$.
The distortion guarantee gives
\begin{equation}
    \|\xv-\xtv\|_2^2\leq nB\delta.
    \label{eq:proof-code-distortion}
\end{equation}
Given $\mathbf{u}\in\mathbb{R}^{nB}$, and $\cv\in\Cc$,  define $\hv(\uv)\triangleq\sum_{i=1}^B\Dv_i\uv_i$ and  $\rv(\cv)\triangleq\hv(\xv)-\hv(\cv)$. That is, for $j=1,\ldots,n$, 
\begin{align}
    h_j(\mathbf{u})
&=\sum_{i=1}^{B}D_{ij}u_{ij},\nonumber\\
    r_j(\cv)&= h_j(\xv)-h_j(\cv),
    \label{eq:proof-residuals}
\end{align}
Also, define
\[
A_T(\cv)
\triangleq
\sqrt{\sum_{j\in\Ic_s}
    \big(T-h_j(\cv)\big)^2
    \ind_{h_j(\cv)\leq T}
+\sum_{j\in\Ic_s^c}r_j(\cv)^2}.
\]
Since $y_{T,j}=T$,  for $j\in\Ic_s$ and
$y_{T,j}=h_j(\xv)+z_j$ for  $j\in \Ic_s^c$, using these definitions, the saturation-aware objective function 
can be written as
\begin{align}
\mathcal{L}_T(\cv)
={}&
\sum_{j\in\Ic_s}
    \big(T-h_j(\cv)\big)^2
    \ind_{h_j(\cv)\leq T}
\nonumber\\[-1mm]
&+
\sum_{j\in\Ic_s^c}
    \big(r_j(\cv)+z_j\big)^2.
\label{eq:proof-loss}
\end{align}
Since $\xhv$ minimizes this objective over all $\cv\in\Cc$  and $\xtv\in\Cc$, we have
\begin{equation}
    \mathcal{L}_T(\xhv)\leq\mathcal{L}_T(\xtv).
    \label{eq:proof-optimality}
\end{equation}
Applying the triangle and reverse-triangle inequalities to
\eqref{eq:proof-optimality}, it follows that
\begin{align}
    A_T(\xhv) &\leq
    A_T(\xtv) +2(\sum_{j\in\Ic_s^c}z_j^2)^{1/2}\nonumber\\
     &\leq
    A_T(\xtv)
    +2\epsilon_z.
    \label{eq:proof-noise}
\end{align}

Note that  $A_T(\xtv)^2 = \sum_{j\in\Ic_s}
    (T-h_j(\xtv))^2
    \ind_{h_j(\xtv)\leq T}
+\sum_{j\in\Ic_s^c}r_j(\xtv)^2$
Since $T-h_j(\xtv)=T-h_j(\xv)+h_j(\xv)-h_j(\xtv)$, we have
\begin{align}
    A_T(\xtv)^2 \leq&  \sum_{j=1}^n r_j(\xtv)^2 +\sum_{j\in\Ic_s}(T-h_j(\xv))^2\nonumber\\
    &+2\sum_{j\in \Ic_s}|T-h_j(\xv)||r_j(\xtv)|\nonumber\\
    \stackrel{\rm (a)}{\leq} & \| \rv (\xtv)\|_2^2 +|\Ic_s|\beta_T(\beta_T+2B\rho),\label{eq:proof-upper-U}
\end{align}
where (a) holds because $|r_j(\xtv)|\leq B\rho$, and for $j\in\Ic_s$, $|T-h_j(\xv)|\leq \beta_T$.

Similarly, $A_T(\xhv)^2 = \sum_{j\in\Ic_s}
    (T-h_j(\xhv))^2
    \ind_{h_j(\xhv)\leq T}
+\sum_{j\in\Ic_s^c}r_j(\xhv)^2$. Therefore,
\begin{align}
A_T(\xhv)^2
\geq{}& \sum_{j\in\Ic_s}
    (T-h_j(\xhv))^2
    \ind_{h_j(\xhv)\leq T}
+\sum_{j\in\Ic_s^c}r_j(\xhv)^2\nonumber\\
=& \sum_{j\in\Ic_s}
    \big((T-h_j(\xv))^2+r_j(\xhv)^2\big)\ind_{h_j(\xhv)\leq T} \nonumber\\
    &+\sum_{j\in\Ic_s} 2r_j(\xhv)(T-h_j(\xv)))\ind_{h_j(\xhv)\leq T}\nonumber\\
&+\sum_{j\in\Ic_s^c}r_j(\xhv)^2\nonumber\\
\geq{}& \|\rv(\xhv)\|_2^2 -2\sum_{j\in\Ic_s}
    |T-h_j(\xv)||r_j(\xhv)|\nonumber\\
    &+\sum_{j\in\Ic_s}
    r_j(\xhv)^2(\ind_{h_j(\xhv)\leq T}-1) \nonumber\\
\geq{}&
\|\rv(\xhv)\|_2^2-|\Ic_s|(2\beta_TB\rho+\beta_T^2),
\label{eq:proof-lower-L}
\end{align}
where the last line follows because $r_j(\xhv)^2(\ind_{h_j(\xhv)\leq T}-1)\geq -\beta_T^2$. 
Combining \eqref{eq:proof-noise}, \eqref{eq:proof-upper-U}, and
\eqref{eq:proof-lower-L} yields
\begin{align}
& \sqrt{\|\rv(\xhv)\|_2^2-|\Ic_s|\beta_T(\beta_T+2B\rho)}\nonumber\\
    &\leq \sqrt{\| \rv (\xtv)\|_2^2 +|\Ic_s|\beta_T(\beta_T+2B\rho)}+2\epsilon_z 
\end{align}
and therefore
\begin{align}
 \|\rv(\xhv)\|_2\leq \| \rv (\xtv)\|_2 +2\sqrt{|\Ic_s|\beta_T(\beta_T+2B\rho)}+2\epsilon_z.
\label{eq:proof-sat-main}
\end{align}
The rest of the proof is similar to the proof of Theorem 4.1 in \cite{zhao2025theoretical}, which we present for completeness. We next employ concentration bounds to connect $\|\rv(\xhv)\|_2$ and $\|\rv(\xtv)\|_2$ to $\|\xhv-\xv\|_2$ and $\|\xtv-\xv\|_2$, respectively. Also, we need to bound $|\Ic_s|$.

For $\cv\in\Cc$, let
\begin{equation}
\begin{aligned}
\Gamma(\cv)
\triangleq{}&
\frac{p^2}{n}
\Big\|
    \sum_{i=1}^{B}(\xv_i-\cv_i)
\Big\|_2^2
+
\frac{p-p^2}{n}\|\xv-\cv\|_2^2.
\end{aligned}
\label{eq:proof-Gamma}
\end{equation}
For any $\epsilon_1,\epsilon_2>0$, define the events $\Ec_1$, $\Ec_2$, and $\Ec_s$ as \begin{align} 
\Ec_1 ={}
& \left\{ \frac{1}{n}\|\rv(\xtv)\|_2^2 \leq \Gamma(\xtv) + \frac{B\rho^2\epsilon_1}{2} \right\}, 
\label{eq:proof-E1}\\ 
\Ec_2 ={}
& \left\{ \frac{1}{n}\|\rv(\cv)\|_2^2 \geq \Gamma(\cv) - \frac{B\rho^2\epsilon_1}{2}, \ \forall\cv\in\Cc \right\}, 
\label{eq:proof-E2}\\ 
\Ec_s ={}
& \left\{ |\Ic_s| \leq \mathbb{E}[|\Ic_s|] + n\epsilon_2 \right\}. 
\label{eq:proof-Es} 
\end{align}

Conditioned on $\Ec_1\cap\Ec_2\cap\Ec_s$, using \eqref{eq:proof-code-distortion} and noting that $\mathbb{E}[|\Ic_s|]=np_s(\xv;T)$, it follows from \eqref{eq:proof-sat-main} that 
\begin{align} 
&\frac{1}{\sqrt{nB}}\|\xv-\xhv\|_2 \leq{} \sqrt{\left(1+\frac{Bp}{1-p}\right)\delta} + 2\rho\sqrt{\frac{\epsilon_1}{2p(1-p)}} \nonumber\\ &+ 2\sqrt{ \frac{ \big(p_s(\xv;T)+\epsilon_2\big) \beta_T(\beta_T+2B\rho) }{ p(1-p)B } } + \frac{2\epsilon_z}{\sqrt{p(1-p)nB}}. 
\end{align}
To finish the proof we need to bound the probability of $(\Ec_1\cap\ldots\cap\Ec_3)^c$. By   the union bound, we have $\P((\Ec_1\cap\Ec_2\cap\Ec_s)^c)
\leq 
\P(\Ec_1^c)+\P(\Ec_2^c)+\P(\Ec_s^c)$.  Using the same analysis performed in \cite{zhao2025theoretical}, it follows that $
    \P(\Ec_1^c)+\P(\Ec_2^c)
\leq
{2^{Br+1}\exp(-\frac{n\epsilon_1^2}{2B^2})}.$
Finally, note that $|\Ic_s|=\sum_{j=1}^{n}S_j$, where
$S_j=\ind_{(\Hv \xv)_j\geq T}$. The variables
$S_1,\ldots,S_n$ are independent Bernoulli random variables. Hence, using the 
Hoeffding's inequality 
\begin{align}
\P(\Ec_s^c)
&\leq
\exp(-2n\epsilon_2^2),
\label{eq:proof-Es-probability}.
\end{align}
which completes the proof.

\subsection{Proof of Corollary~\ref{cor:1}}

Since $\epsilon_z=0$, set
\begin{equation}
\begin{aligned}
&\delta'=\frac{\delta}{\rho^2}~,~~~
T'=\frac{T}{\rho},\\
\text{and}\\
&\Delta_T=\left(\frac{B}{2}-T'\right)^+
\left[
    \left(\frac{B}{2}-T'\right)^+ +2B
\right].
\end{aligned}
\label{eq:proof-normalized-parameters}
\end{equation}
The quantity $\Delta_T$ does not depend on $p$. Minimizing the recovery
bound is equivalent to minimizing
\begin{align}
&g(p;T) ={} \left(1+\frac{Bp}{1-p}\right)\delta' + \frac{\epsilon_1}{p(1-p)} \nonumber\\ 
&+ \frac{ 2\big(p_s(\xv;T)+\epsilon_2\big)\Delta_T }{ Bp(1-p) }.
\label{eq:proof-g}
\end{align}
Its derivative is
\begin{equation}
\begin{aligned}
&g'(p;T) ={} \frac{B\delta'}{(1-p)^2} + \frac{\epsilon_1(2p-1)} {p^2(1-p)^2} \nonumber\\ 
&+ \frac{2\Delta_T}{B} \left[ \frac{1}{p(1-p)} \frac{\partial p_s(p,T)}{\partial p} + \frac{ \big(p_s(p,T)+\epsilon_2\big)(2p-1) }{ p^2(1-p)^2 } \right].
\end{aligned}
\label{eq:proof-g-prime}
\end{equation}
Let $p_T^*$ satisfy $g'(p_T^*;T)=0$. Rearranging gives
\begin{equation}
\begin{aligned}
&(1-2p_T^*) \left[ \epsilon_1 + \frac{2\Delta_T}{B} \big(p_s(p_T^*,T)+\epsilon_2\big) \right] \nonumber\\ &\qquad = B(p_T^*)^2\delta' + \frac{2\Delta_T}{B} p_T^*(1-p_T^*) \left. \frac{\partial p_s(p,T)}{\partial p} \right|_{p=p_T^*}.
\end{aligned}
\label{eq:proof-corollary-stationarity}
\end{equation}
Increasing $p$ increases the number of nonzero mask entries and hence
$p_s(\xv;T)$. Thus, $\partial p_s(\xv;T)/\partial p\geq0$, so the
right-hand side of \eqref{eq:proof-corollary-stationarity} is positive.
It follows that $p_T^*<1/2$.

To establish the dependence on $T$, define 
\[
\Psi(p,T)
=
\frac{\Delta_T\bigl(p_s(p,T)+\epsilon_2\bigr)}{p(1-p)}.
\]
The derivative $\Psi_p(p,T)$ is nonincreasing in $T$, while
$g(p;T)$ is convex in $p$. For $T_1<T_2$, using
$g'(p_{T_1}^*;T_1)=0$, we obtain
\begin{align} 
&g'(p_{T_1}^*;T_2) = g'(p_{T_1}^*;T_1) \\\nonumber
&+ \frac{2}{B} \left[ \Psi_p(p_{T_1}^*,T_2) - \Psi_p(p_{T_1}^*,T_1) \right] \leq 0. 
\end{align} 

Since $g'(p;T_2)$ is increasing in $p$, its unique zero satisfies $p_{T_2}^*\geq p_{T_1}^*$. Therefore, $p_T^*$ is nondecreasing in $T$.



\section{Conclusion}
In this paper, we presented the first systematic analysis of SCI systems under measurement saturation. Our theoretical results establish that binary masks with Bernoulli distribution achieve optimal recovery when the probability of ones ($p$) is below $0.5$, and that stronger saturation requires even smaller $p$. These insights provide a principled guideline for mask design in practice. Complementing the theory, we proposed SAPnet, a saturation-aware plug-and-play reconstruction algorithm that introduces a consistency loss term to explicitly account for clipped measurements. Experiments on standard video SCI benchmarks validate the theory and demonstrate substantial PSNR improvements over PnP-FastDVDnet, while maintaining low computational cost and robustness across saturation levels and mask patterns. 

\bibliographystyle{IEEEbib}
\bibliography{myrefs.bib}


\end{document}